\documentclass[conference]{IEEEtran}
\IEEEoverridecommandlockouts
\usepackage{cite}
\usepackage{amsmath,amssymb,amsfonts}
\usepackage{algorithmic}
\usepackage{graphicx}
\usepackage{textcomp}
\usepackage{xcolor}
\usepackage{listings}
\usepackage{tikz}
\usetikzlibrary{arrows.meta,positioning,fit,calc,backgrounds}
\usepackage[hyphens]{url}
\makeatletter
\newcommand{\linebreakand}{%
  \end{@IEEEauthorhalign}
  \hfill\mbox{}\par
  \mbox{\hspace{0.5em}\null}\hfill\begin{@IEEEauthorhalign}
}
\makeatother

\begin{document}

\title{JetSCI: A Hybrid JAX–PETSc Framework for Scalable Differentiable Simulation\\
\thanks{This work is funded under the U.S. Air Force Research Laboratory contract Aerospace Materials Processing, Performance and Characterization (AMPPAC), FA2394-25-D-B005.}
}

\author{\IEEEauthorblockN{Alberto Cattaneo}
\IEEEauthorblockA{\textit{Kahlert School of Computing} \\
\textit{University of Utah}\\
Salt Lake City, Utah, USA \\
u1472208@utah.edu}
\and
\IEEEauthorblockN{M. Keith Ballard}
\IEEEauthorblockA{\textit{Materials and Manufacturing Directorate} \\
\textit{Composite Performance, Air Force Research Laboratory}\\
Wright-Patterson AFB, Ohio, USA \\
michael.ballard.28@afrl.af.mil}
\linebreakand
\IEEEauthorblockN{Robert M. Kirby}
\IEEEauthorblockA{\textit{Scientific Computing and Imaging Institute} \\
\textit{University of Utah}\\
Salt Lake City, Utah, USA \\
Mike.Kirby@utah.edu}
\and
\IEEEauthorblockN{Varun Shankar}
\IEEEauthorblockA{\textit{Kahlert School of Computing} \\
\textit{University of Utah}\\
Salt Lake City, Utah, USA \\
shankar@cs.utah.edu}
}

\maketitle
\begin{abstract}
The rapid rise of scientific machine learning (SciML) has expanded the role of differentiable modeling, surrogate modeling, and data-driven constitutive laws in large-scale simulation. The JAX framework provides an attractive environment for these workflows through automatically differentiable programs, vectorization, GPU acceleration, and while enabling seamless learning of surrogate models. However, large-scale simulation still relies on mature HPC infrastructure. Libraries, such as PETSc, provide scalable MPI-based parallelism, robust linear and nonlinear solvers, and advanced preconditioning capabilities that remain difficult to reproduce in JAX-only workflows. We present JetSCI, a hybrid JAX–PETSc framework that unifies these complementary strengths. JetSCI uses JAX for GPU-parallel differentiable discretizations and PETSc for robust, scalable solution of the resulting systems on distributed-memory architectures, exposing multilevel parallelism through GPU acceleration within nodes and MPI parallelism across nodes. For finite element discretizations of heterogeneous micromechanics problems, JetSCI outperforms JAX-only implementations in efficiency and accuracy.
\end{abstract}


\section{Introduction}
\label{sec:intro}
Scientific machine learning (SciML) is changing how large-scale simulation codes are constructed, coupled, and deployed. Across computational science, machine learning is no longer used only for post-processing or reduced-order closure; it is increasingly embedded directly into simulation workflows through differentiable discretizations, learned surrogate models, operator-learning architectures, and data-driven constitutive updates \cite{karniadakis2021physics,subedi2026operator,herrmann2024deep,fuhg2025review}. In computational mechanics in particular, this shift has created new opportunities to combine first-principles models with learned constitutive behavior and multiscale surrogates, while also exposing new requirements for software systems that can support automatic differentiation, accelerator execution, and scalable PDE solves within a single workflow \cite{herrmann2024deep,upadhyay2024physics,fuhg2025review}.

Modern array-programming frameworks have made this style of simulation increasingly practical. The scientific Python ecosystem has become a central interface for numerical computing, with NumPy providing the dominant array abstraction and SciPy supplying much of the core algorithmic infrastructure used throughout scientific and engineering software \cite{harris2020numpy,virtanen2020scipy}. Built on this ecosystem, JAX provides composable transformations for numerical programs, including automatic differentiation, vectorization, and just-in-time compilation, together with execution on accelerator hardware \cite{bradbury2018jax}. These features make JAX attractive for differentiable simulation, inverse problems, and machine-learning-augmented numerical methods, especially when local residual evaluations, constitutive updates, and element-level kernels can be expressed in an array-oriented style. This promise is already visible in emerging differentiable mechanics frameworks such as JAX-FEM, which show that substantial portions of finite-element workflows can be written concisely while retaining access to automatic differentiation and GPU-backed execution \cite{xue2023jaxfem}.

However, the strengths of JAX do not remove the core algorithmic and software demands of large-scale PDE simulation. Once problem sizes become sufficiently large, or when implicit time stepping, nonlinear constitutive response, multiphysics coupling, or highly heterogeneous coefficients are present, end-to-end performance is often governed less by local kernel throughput than by the global solution of large sparse algebraic systems. In this regime, scalable distributed-memory data structures, robust Krylov solvers, nonlinear solution strategies, preconditioning, and matrix-free formulations are not optional implementation details but central algorithmic components. These capabilities have been developed over decades in production HPC libraries rather than in ML-oriented frameworks \cite{balay1997efficient,petsc_users_manual}. As a result, a purely JAX-based implementation may offer elegant differentiation and accelerator programming while still lacking the mature distributed solver stack needed for robust simulation at scale.

PETSc is one of the standard foundations for this solver-centric view of scientific computing. Designed for the scalable solution of PDE-based applications, PETSc provides distributed vectors and matrices, Krylov subspace methods, nonlinear solvers, preconditioners, time integrators, optimization tools, and support for both matrix-based and matrix-free workflows, all organized around MPI-based distributed-memory execution \cite{balay1997efficient,petsc_users_manual,petsc-web-page}. It has consequently become a common substrate for large-scale multiphysics and multiscale simulation codes in which solver robustness, portability, and parallel scalability are decisive. Yet PETSc, like most mature HPC libraries, is not designed around differentiable programming or close integration with modern machine-learning frameworks. This creates a persistent software and algorithmic gap: ML-oriented systems provide expressive differentiable kernels and convenient accelerator programming, whereas solver-oriented HPC systems provide the distributed numerical infrastructure needed once simulation becomes dominated by sparse global solves.

The central idea of this work is that these two ecosystems should not be viewed as competitors, but as complementary layers in a single computational stack. We present \emph{JetSCI}, a hybrid JAX--PETSc framework that combines JAX for differentiable, accelerator-oriented discretization and constitutive computation with PETSc for robust and scalable distributed-memory solution of the resulting algebraic systems. This separation of concerns is natural from both a software and numerical perspective: local physics, automatic differentiation, and data-driven updates are expressed where JAX is strongest, while global linear and nonlinear solution is delegated to PETSc, where decades of solver development can be brought to bear. The resulting design exposes multilevel parallelism through accelerator execution within nodes and MPI parallelism across nodes, while avoiding the need to reproduce mature sparse-solver technology inside a JAX-only workflow.

This hybridization is especially relevant for computational mechanics problems involving heterogeneous and multiscale material behavior. Such problems are natural targets for learned constitutive laws, differentiable calibration, and surrogate-assisted multiscale simulation, but they also generate stiff, ill-conditioned, and communication-intensive algebraic systems whose efficient solution still depends on established HPC solver infrastructure \cite{herrmann2024deep,fuhg2025review,upadhyay2024physics}. JetSCI is designed precisely for this setting. In the numerical experiments reported here, we consider finite-element discretizations of heterogeneous micromechanics problems and show that coupling JAX and PETSc yields a practical framework that improves efficiency, robustness, and scalability relative to JAX-only approaches, while preserving the differentiable programming model needed for modern SciML workflows.\cite{chatgpt2026}

\subsection{Related Work}

Differentiable programming for PDE-constrained simulation now spans several distinct software design points. At one end are \emph{ML-first} differentiable simulation frameworks, which implement both local discretization kernels and global solves largely inside an automatic-differentiation-enabled array framework. \textbf{JAX-FEM}~\cite{xue2023jaxfem} is the canonical example in computational mechanics: it uses JAX to provide differentiable, GPU-accelerated finite-element workflows for forward and inverse problems. More recent JAX-based efforts such as \textbf{JAX-SSO}~\cite{wu2024jaxsso} pursue a similar direction for structural optimization and neural-network coupling. Earlier TensorFlow-based efforts such as \textbf{ADCME}~\cite{xu2020adcme} likewise demonstrated the appeal of embedding physical simulation and inverse problems in a modern AD framework. These systems provide seamless end-to-end differentiability and easy integration with machine learning models, but they largely retain both discretization and solver logic inside a single ML-oriented runtime. JetSCI shares the interest in ML-native local differentiation, but deliberately does not require the global sparse solve to remain inside JAX.

A second line of work is represented by high-level finite-element domain-specific languages (DSLs) and adjoint-capable partial differential equation (PDE) frameworks. \textbf{Firedrake}~\cite{Firedrake} and \textbf{FEniCSx/DOLFINx}~\cite{BarattaEtal2023} provide mature finite-element environments with MPI-parallel execution and PETSc-based linear and nonlinear solver support. In the Firedrake/FEniCS ecosystem, \textbf{dolfin-adjoint}/\textbf{pyadjoint} automates discrete adjoints for PDE-constrained optimization and inverse problems~\cite{mitusch2019dolfinadjoint}. Recent work has extended this direction by coupling Firedrake directly to PyTorch and, more generally, by introducing differentiable-programming abstractions that connect Firedrake-based PDE solvers to PyTorch and JAX models~\cite{bouziani2023physicsdriven,bouziani2024barrier}. Ongoing work in the FEniCSx ecosystem, such as \textbf{FormOpt}, continues this emphasis on scalable sensitivity-based optimization on top of a PETSc-backed FE framework~\cite{diazavalos2026formopt}. These systems are highly expressive and productive, but their differentiation mechanisms are centered on variational forms, generated adjoints, and symbolic operator composition. JetSCI instead works one level lower in the stack: it uses JAX directly for element-level residual and tangent generation and hands explicit sparse matrices and vectors to PETSc, keeping the solver layer fully visible.

A third and increasingly important direction augments \emph{existing HPC codes} with AD rather than rewriting them inside an ML framework. The most prominent recent example is \textbf{Enzyme}, which performs automatic differentiation on optimized LLVM IR and has been extended to GPU kernels and multiple parallel programming paradigms~\cite{moses2020enzyme,moses2021enzymegpu,moses2022parallelenzyme}. This compiler-level route is especially attractive for large legacy codes, since it can preserve native data structures, languages, and solver stacks. Recent \textbf{MFEM} work pushes this idea directly into large-scale finite elements by localizing AD at the finite-element-operator-decomposition level, so that local dense derivatives are composed with globally sparse FE operators in a scalable C++ workflow~\cite{andrej2025scalablead}. Likewise, the broader C++ finite-element ecosystem has long supported AD through operator-overloading libraries and templated residual evaluations: \textbf{deal.II} supports AD backends such as \textbf{ADOL-C} and \textbf{Sacado}~\cite{arndt2021dealii,adolc1996,phipps2022sacado}, \textbf{MOOSE} provides AD via \textbf{MetaPhyicL}~\cite{harbour2025moose,lindsay2021automatic}, and \textbf{lifeX} explicitly exposes AD support in a PETSc-backed HPC FE library built on deal.II~\cite{africa2022lifex}. Related high-performance AD packages such as \textbf{CoDiPack} continue this trend of bringing efficient derivative computation to native C++ simulation codes~\cite{sagebaum2019codipack}. JetSCI is philosophically close to this family in that it respects finite-element locality and avoids dense global AD, but it makes a different trade-off: instead of compiler-level AD inside a monolithic native codebase, it uses JAX as a high-productivity frontend for local differentiated kernels and PETSc as the backend for the resulting sparse global algebra.

There are also lower-level interoperability efforts aimed at composing external HPC kernels with ML frameworks. \textbf{JAXbind}, for example, reduces the effort required to bind external functions into JAX, but currently focuses on CPU-side bindings and requires the user to supply Jacobian-product rules for wrapped primitives~\cite{roth2024jaxbind}. More generally, Enzyme-based bindings provide a route for exposing foreign LLVM/MLIR code to JAX-style workflows~\cite{moses2020enzyme,roth2024jaxbind}. On the solver side, \textbf{petsc4py}~\cite{DALCIN20111124} exposes PETSc cleanly in Python and is a key enabling layer for JetSCI, but by itself it provides neither automatic differentiation nor a JAX-native element-kernel workflow. JetSCI therefore builds on this interop substrate in a more structured way: JAX is used exactly where automatic differentiation and accelerator execution are most effective, while PETSc remains the owner of the sparse algebraic objects and distributed-memory solves.

Overall, JetSCI occupies a middle ground between ML-first differentiable simulators and AD-retrofitted native HPC codes. Relative to pure-JAX approaches, it gives up a fully single-runtime workflow in exchange for direct access to PETSc's mature sparse solver, preconditioning, and MPI infrastructure. Relative to Enzyme-, Sacado-, or ADOL-C-based native C++ approaches, it gives up some low-level uniformity in exchange for a Pythonic, accelerator-friendly, ML-native environment for local constitutive and assembly kernels. We view this separation as a strength rather than a compromise: in the finite-element setting \texttt{local dense differentiated kernels} $\to$ \texttt{global sparse assembly} $\to$ \texttt{distributed solve},
JetSCI allows each stage to be implemented in the software layer best matched to its numerical and performance requirements.\cite{chatgpt2026}
\section{Finite-element and software background}

\subsection{Finite-element problem structure}
We consider nonlinear finite-element simulations of heterogeneous materials and microstructured solids, for which the discrete problem takes the residual form
\[
R(u)=0,
\]
where $R(u)$ is a nonlinear function of the nodal displacement vector $u$. JetSCI solves this with Newton's method. Newton's method in turn requires the global Jacobian operator
\[
K(u)=\frac{\partial R}{\partial u}.
\]
The global residual and Jacobian are assembled from element-level contributions. The assembled global Jacobian is sparse, reflecting the locality of the finite-element discretization, whereas the element-level Jacobians are small and dense as a consequence of per-element polynomial and constitutive computations. The sparse Jacobian system is linear, requiring efficient linear solvers within the outer Newton iterations. This separation of scales is central to JetSCI: local residual and Jacobian evaluation are natural targets for accelerator-backed automatic differentiation, while the resulting global sparse systems still require robust and scalable linear and nonlinear solvers.

\subsection{JAX and \texttt{fea-in-jax}}
On the discretization side, JetSCI builds on \texttt{fea-in-jax}, a JAX-based finite-element implementation developed by Keith Ballard~\cite{ballard_feainjax}. JAX provides a NumPy-like programming model together with automatic differentiation, just-in-time compilation, and vectorization transformations, making it well suited for repeated element-batch computations on accelerators~\cite{bradbury2018jax,jax_docs}. The \texttt{fea-in-jax} codebase provides the JAX-side FEM substrate used here, including residual-based Newton solves, Jacobian formation, and treatment of boundary conditions. In JetSCI, this layer is retained for local finite-element computation, differentiated kernel evaluation, and the Newton solve, while the principal extensions are explicit sparse matrix assembly, low-overhead PETSc handoff, and distributed-memory execution.

This starting point is particularly natural for the heterogeneous micromechanics problems considered in this work~\cite{ballard2014microstructure,ballard2017tow,ballard2018textile}, where complex local constitutive and geometric information must be propagated into large sparse algebraic systems. JetSCI preserves the flexibility of a JAX-based element formulation for these local kernels, but couples it to PETSc so that the assembled systems can be solved with mature sparse solver, preconditioning, and MPI infrastructure.

\subsection{PETSc, \texttt{petsc4py}, and \texttt{mpi4py}}
PETSc provides the scalable solver backbone for JetSCI; in this work, we leverage PETSc's linear solvers and preconditioners within the \texttt{fea-in-jax} Newton solve. It includes parallel matrix and vector assembly routines together with linear and nonlinear solvers, preconditioners, and support for CPU and GPU execution on distributed-memory systems~\cite{petsc_users_manual,petsc_overview}. Rather than interacting with PETSc through a high-level finite-element DSL, JetSCI uses \texttt{petsc4py}, the Python interface to PETSc, so that PETSc objects and solver configurations remain explicit at the application level~\cite{dalcin_petsc4py,petsc4py_docs}. Distributed-memory execution is exposed through MPI, and JetSCI relies on \texttt{mpi4py} for Python-level access to MPI semantics and rank-local program structure~\cite{dalcin2011python,dalcin2021mpi4py,mpi4py_docs}. This combination allows the JAX side of the framework to remain focused on local differentiated kernels while PETSc owns the globally distributed sparse algebraic objects and their associated solves.

\subsection{CuPy, DLPack, and \texttt{ctypes}}
A key implementation challenge in JetSCI is transferring accelerator-resident matrix data from the JAX side into PETSc without unnecessary host staging. For this purpose, JetSCI uses \texttt{CuPy} as a lightweight GPU-array interoperability layer. CuPy provides a NumPy/SciPy-compatible array interface on GPU devices and exposes device pointers in a form convenient for Python-side systems integration~\cite{okuta2017cupy,cupy_docs}. To move data between JAX-managed arrays and CuPy without materializing new host copies, JetSCI uses \texttt{DLPack}, which defines a standard in-memory tensor interchange mechanism across array frameworks~\cite{dlpack_spec,dlpack_repo}. However, \texttt{petsc4py} does not by itself expose the full PETSc path for constructing GPU-resident sparse matrices directly from such device buffers. JetSCI therefore uses Python's \texttt{ctypes} foreign-function interface to call the relevant PETSc C routine directly from Python, passing device pointers obtained from the CuPy view of the JAX data~\cite{python_ctypes}. In this way, CuPy and DLPack provide the device-array exchange layer, while \texttt{ctypes} provides the final low-level bridge into PETSc's native matrix-construction interface.

\section{The JetSCI Framework}
\label{sec:jetsci}
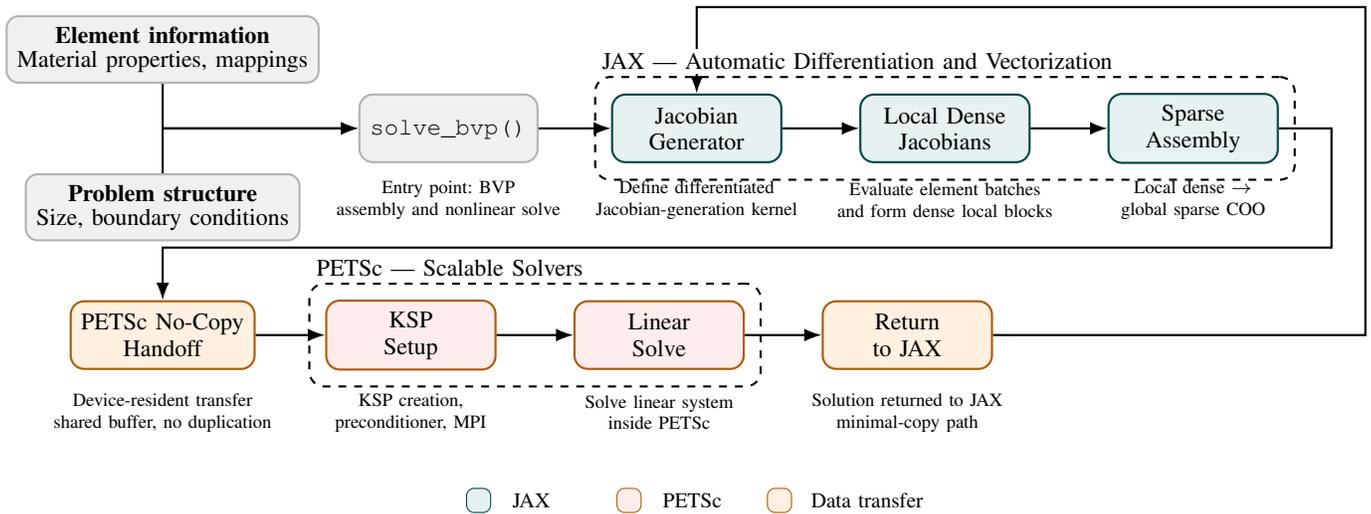
\begin{figure*}[t]
\centering
\begin{tikzpicture}[
    font=\small,
    >=Latex,
    stage/.style={
        draw,
        rounded corners=1.8mm,
        minimum width=2.25cm,
        minimum height=0.90cm,
        align=center,
        thick,
        inner sep=4pt
    },
    inputbox/.style={
        draw,
        rounded corners=1.6mm,
        minimum width=2.55cm,
        minimum height=0.86cm,
        align=center,
        thick,
        inner sep=4pt,
        fill=gray!12,
        draw=gray!60
    },
    neutral/.style={stage, fill=gray!12, draw=gray!60},
    jax/.style={stage, fill=teal!10, draw=teal!55!black},
    transfer/.style={stage, fill=orange!12, draw=orange!70!black},
    petsc/.style={stage, fill=red!7, draw=orange!70!black},
    group/.style={
        draw,
        dashed,
        rounded corners=2mm,
        inner sep=6pt,
        thick
    },
    note/.style={
        font=\scriptsize,
        text=black,
        align=center
    },
    grouplabel/.style={
        font=\small,
        fill=white,
        inner sep=1.5pt,
        align=left
    },
    legendbox/.style={
        draw,
        rounded corners=1mm,
        minimum width=0.32cm,
        minimum height=0.32cm
    }
]

\node[inputbox] (elem) at (1.8,1.05) {
    \textbf{Element information}\\[-1pt]
    Material properties, mappings
};

\node[inputbox] (prob) at (1.8,-1.05) {
    \textbf{Problem structure}\\[-1pt]
    Size, boundary conditions
};


\node[neutral] (solve)     at (5.6,0.0)  {\texttt{solve\_bvp()}};
\node[jax]     (jacgen)    at (8.9,0.0)  {Jacobian\\Generator};
\node[jax]     (localjac)  at (12.2,0.0) {Local Dense\\Jacobians};
\node[jax]     (sparseasm) at (15.5,0.0) {Sparse\\Assembly};

\draw[->, thick, black] (elem.south) |- (solve.west);
\draw[->, thick, black] (prob.north) |- (solve.west);

\draw[->, thick, black] (solve.east) -- (jacgen.west);
\draw[->, thick, black] (jacgen.east) -- (localjac.west);
\draw[->, thick, black] (localjac.east) -- (sparseasm.west);

\node[transfer] (handoff)  at (1.8,-2.75)  {PETSc\ No-Copy\\ Handoff};
\node[petsc]    (ksp)      at (5.1,-2.75)  {KSP\\Setup};
\node[petsc]    (linsolve) at (8.4,-2.75)  {Linear\\Solve};
\node[transfer] (retjax)   at (11.7,-2.75) {Return\\to JAX};


\coordinate (route1) at ($(sparseasm.east)+(0.70,0)$);
\coordinate (handoffcap) at ($(handoff.north)+(0,0.70)$);
\draw[->, thick, black] (sparseasm.east) -- (route1) |- (handoffcap) -- (handoff.north);

\draw[->, thick, black] (handoff.east) -- (ksp.west);
\draw[->, thick, black] (ksp.east) -- (linsolve.west);
\draw[->, thick, black] (linsolve.east) -- (retjax.west);
\coordinate (fb1) at ($(retjax.east)+(4.95,0)$);
\coordinate (fb2) at ($(fb1 |- jacgen.north)+(0,1.15)$);
\draw[->, thick, black] (retjax.east) -- (fb1) |- (fb2) -| (jacgen.north);
\node[group, fit=(jacgen)(localjac)(sparseasm)] (jaxgroup) {};
\node[group, fit=(ksp)(linsolve)] (petscgroup) {};

\node[grouplabel, anchor=west] at ($(jaxgroup.north west)+(0.05,0.20)$)
    {JAX --- Automatic Differentiation and Vectorization};

\node[grouplabel, anchor=west] at ($(petscgroup.north west)+(0.05,0.20)$)
    {PETSc --- Scalable Solvers};


\node[note] at (5.6,-0.95)  {Entry point: BVP\\assembly and nonlinear solve};
\node[note] at (8.9,-0.95)  {Define differentiated\\Jacobian-generation kernel};
\node[note] at (12.2,-0.95) {Evaluate element batches\\and form dense local blocks};
\node[note] at (15.5,-0.95) {Local dense $\rightarrow$\\global sparse COO};

\node[note] at (1.8,-3.78)  {Device-resident transfer\\shared buffer, no duplication};
\node[note] at (5.1,-3.78)  {KSP creation,\\preconditioner, MPI};
\node[note] at (8.4,-3.78)  {Solve linear system\\inside PETSc};
\node[note] at (11.7,-3.78) {Solution returned to JAX\\minimal-copy path};

\node[legendbox, fill=teal!10, draw=teal!55!black] (leg1) at (6,-4.95) {};
\node[right=0.16cm of leg1, anchor=west, font=\footnotesize] {JAX};

\node[legendbox, fill=red!7, draw=orange!70!black] (leg2) at (8.0,-4.95) {};
\node[right=0.16cm of leg2, anchor=west, font=\footnotesize] {PETSc};

\node[legendbox, fill=orange!12, draw=orange!70!black] (leg3) at (10,-4.95) {};
\node[right=0.13cm of leg3, anchor=west, font=\footnotesize] {Data transfer};

\end{tikzpicture}
\caption{JetSCI workflow. Element information and global problem structure both feed into \texttt{solve\_bvp()}. The JAX FEM layer, built on \texttt{fea-in-jax}, performs nonlinear solves using Newton's method, defines differentiated Jacobian-generation kernels based on defined residual kernels, evaluates element batches to form dense local Jacobian contributions, and assembles global sparse COO matrices. These device-resident matrix entries are then handed to PETSc for KSP setup and linear solution, after which the solution is returned to JAX.}
\label{fig:jetsci_flow}
\end{figure*}
JetSCI is a hybrid JAX--PETSc framework for differentiable, distributed-memory simulation. Its design separates local differentiated computation from global algebraic solution. JAX is used where automatic differentiation, accelerator-backed array programming, and JIT-compiled element-level kernels are most valuable, while PETSc is used where mature sparse linear algebra, preconditioning, nonlinear solvers, and MPI-based distributed-memory execution are essential. This division of labor is the central design principle of JetSCI. On the JAX side, JetSCI builds on \texttt{fea-in-jax}, an existing JAX-based finite-element implementation, and extends it with explicit sparse matrix construction, PETSc solver integration, low-overhead device-resident matrix transfer, and MPI-distributed execution \cite{ballard_feainjax}. The full framework is visualized in Figure \ref{fig:jetsci_flow}.

\subsection{JetSCI design and matrix-explicit integration}
\label{sec:jetsci-design}
The complementary designs of JAX and PETSc admit several integration strategies, each with distinct trade-offs in performance, memory usage, and implementation complexity. The first is a \textit{matrix-free} approach, in which PETSc's matrix shell (\texttt{Mat SHELL}) mechanism is used to define a linear operator whose matrix-vector product is evaluated by a JAX function. In this case, JAX is responsible only for computing the action of the operator on a given vector, while PETSc drives the overall solver iteration. This avoids the cost of explicitly assembling a global matrix and can be memory-efficient for large problems, but it restricts the available preconditioners to those that do not require an assembled matrix and introduces overhead at the JAX--PETSc boundary on each solver iteration.

The second is a \textit{matrix-explicit} approach, in which JAX is used to construct the sparse Jacobian matrix via automatic differentiation, which is then transferred into a PETSc matrix object and handed off entirely to PETSc for solving. This incurs a one-time assembly cost and a larger memory footprint, but grants access to PETSc's full suite of preconditioners and solvers, eliminates repeated framework boundary crossings during the solve, and enables more robust convergence on ill-conditioned systems. Having tested both options, we found that the overhead incurred in the matrix shell paradigm due to passing back and forth between frameworks overwhelms any benefits. Consequently, JetSCI adopts the matrix-explicit approach: we create the relevant data in JAX, transfer it to PETSc, perform all solves within that framework, and then pass data back as needed. Since JetSCI focuses extensively on matrix construction and transfer, we discuss these details below.

\subsection{JAX matrix creation}
\label{sec:jetsci-mat}
While there is a higher memory cost associated with explicit matrix construction, it can be lessened via the use of sparse matrices wherever possible. As such, JetSCI utilizes a batch-based system for creating the Jacobian. Each batch consists of elements that have a similar constitutive model, number of nodes, order of basis functions, and quadrature, producing arrays of the same shape. By doing this we can JIT-compile the handling of each batch of elements efficiently, since JAX requires consistency of shape and operation. A pseudocode version of this process is shown below.

\begin{lstlisting}
nnz = PrecomputeNonZeros  # needed to allow jitting

@jax.vmap
def generateJacobian(element):
    return jax.jacfwd(element.residual_kernel)

for b in batch:
    local_dense_jacobian_values = generateJacobian(b.elements)
    global_indices = b.global_component_index
    values.append(local_dense_jacobian_values)
    indices.append(global_indices)

data, rows, cols = sortAndRemoveDuplicate(values, indices)
sparse_COO = assembleCOOTuple(data, rows, cols, nnz)
\end{lstlisting}

This matrix, represented in sparse format, is passed into a PETSc object, at which point all further global algebraic computations are performed by PETSc. This cuts out overhead from repeated switching between frameworks, while also reducing the number of necessary copies. Furthermore, because it is represented as a PETSc object, it is automatically amenable to PETSc transformations.

The creation of the matrix is handled entirely via JAX's automatic differentiation. Instead of creating a matrix entry by entry, we define a function that performs the appropriate Jacobian-vector product; this function is then automatically differentiated to generate a new function that creates the Jacobian matrix for a given input. Because this is performed automatically, the generating function is optimized for the available hardware, compiled for improved performance on subsequent calls, and is less likely to suffer from human error. This matrix is then passed to PETSc. While in this paper we use an explicitly designed material property function relating stresses and strains, using a surrogate model is equally possible, and the same general steps would be used to extract a Jacobian from such a model.

\subsection{Matrix and vector transfer}
\label{sec:jetsci-transfer}
One practical limitation of working within the PETSc4py framework is that, unlike PETSc's native C interface, PETSc4py does not natively support constructing matrices directly from GPU-resident data. As a result, the default matrix creation path requires the assembled COO data to be transferred off the GPU and back, introducing a redundant device round-trip. While JAX and PETSc both employ sophisticated memory transfer protocols that keep this overhead modest relative to the overall solve time, it nonetheless represents an unnecessary cost, particularly as problem sizes grow and matrix assembly becomes more frequent.

There are workarounds to this. Because the underlying PETSc method can operate using only device data, it is possible to directly call that function after converting the Python object to a pointer. This costs much of the convenience of using PETSc4py's native constructors, but it does decrease the overhead. To rectify this, JetSCI provides a pipeline in which a JAX array created on the GPU exposes a pointer to its device data, which is then extracted and handed to the PETSc matrix creation method exposed via \texttt{ctypes}. An example of this extraction and conversion is shown below.

Because JAX does not expose pointers, it is necessary to use DLPack to convert the JAX object into a format which does. Without this, it is impossible to perform no-copy data movement, and in fact PETSc4py will, by default, stage the arrays on the CPU even if all data used to create the PETSc object is already on the GPU. This holds true for returning the PETSc object data back into a JAX object without staging on the GPU. 

\begin{lstlisting}
def buildPETSc(jacMatVals, jacMatRows, jacMatCols, jacMatShape):
    mat = PETSc.Mat().create(comm=comm)
    mat.setSizes(jacMatShape)
    mat.setType(PETSc.Mat.Type.AIJCUSPARSE)
    mat.setPreallocationCOO(jacMatRows, jacMatCols)

    GPUPointerArray = cp.from_dlpack(jacMatVals, copy=False)

    lib = ct.CDLL(PETSc.file)

    MatSetValuesCOO = lib.MatSetValuesCOO
    MatSetValuesCOO.restype = ct.c_int
    MatSetValuesCOO.argtypes = [ct.c_void_p, ct.c_void_p, ct.c_int]
    mat_ptr = ct.c_void_p(mat.handle)
    coo_ptr = ct.c_void_p(GPUPointerArray.data.ptr)

    MatSetValuesCOO(mat_ptr, coo_ptr, PETSc.InsertMode.INSERT_ALL)
\end{lstlisting}
This avoids expensive \texttt{GPU} $\to$ \texttt{CPU} $\to$ \texttt{GPU} transfers as well as data duplication, thereby decreasing overhead. Once converted, PETSc4py can still be used to handle PETSc solver usage from within a Python paradigm. To avoid breaking JAX's ``no-side-effects'' rule, callbacks are used to create these PETSc objects outside of JAX's scope, and they are called outside of the JAX ecosystem.


\subsection{MPI support}
\label{sec:jetsci-mpi}
While JAX simplifies intra-node CUDA parallelism and GPU memory management, many large-scale simulation workloads require a second level of parallelism: distributed-memory execution across multiple nodes. JetSCI exposes this capability by incorporating MPI through PETSc's parallel infrastructure. JAX does provide its own mechanism for multi-device parallelism through sharding, in which arrays are partitioned across devices according to a user-specified layout and collective operations are handled automatically by the XLA runtime. Sharding is highly effective for structured, dense array computations with regular communication patterns, the kinds of operations common in deep learning workloads. However, scientific computing programs frequently involve unstructured meshes, irregular sparsity patterns, and asynchronous inter-process communication that do not conform to the structured partitioning model sharding requires. Assembly of sparse global operators, halo exchanges on unstructured grids, and adaptive solver communication patterns are difficult or impossible to express cleanly within JAX's sharding abstractions. For these workloads, MPI-based distribution remains the more appropriate and flexible paradigm.

Incorporating MPI alongside JAX and PETSc introduces additional complexity, as data must be shared not only across frameworks but also across devices, which must then be synchronized manually. The use of direct pointer-based, no-copy object creation reduces the safety guarantees that each individual library can independently enforce, requiring careful coordination at framework boundaries. The inclusion of MPI also expands the software stack, as additional libraries must be linked and configured correctly. Despite this added complexity, the integration is made manageable by the fact that PETSc handles all synchronization of MPI-distributed objects once they have been created. The primary responsibility of JetSCI in this context is therefore to translate JAX-side objects into PETSc-compatible formats and to structure JAX computations so that each MPI rank operates only on its local portion of the problem.

While JAX arrays for local computations must be instantiated on each rank in the current implementation, the assembled PETSc matrices and vectors are fully distributed and never replicated globally. One current limitation is that this scheme cannot maintain a strict zero-copy data path: constructing local PETSc objects from globally-aware JAX quantities requires a transfer and therefore a copy. However, for multi-GPU systems where memory capacity is the primary constraint, the cost of this copy is negligible compared to the alternative of failing to fit the problem in memory at all.

\section{Implementation Details}
\label{sec:impl}

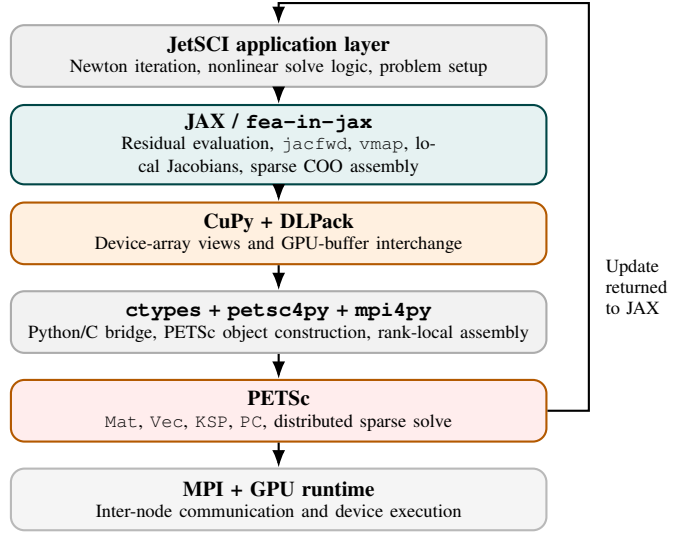
\begin{figure}[t]
\centering
\begin{tikzpicture}[
    >=Latex,
    font=\footnotesize,
    layer/.style={
        draw,
        rounded corners=1.8mm,
        minimum width=0.80\columnwidth,
        text width=0.78\columnwidth,
        minimum height=0.82cm,
        align=center,
        thick,
        inner sep=3pt
    },
    top/.style={layer, fill=gray!12, draw=gray!65},
    jax/.style={layer, fill=teal!10, draw=teal!55!black},
    interop/.style={layer, fill=orange!12, draw=orange!70!black},
    bridge/.style={layer, fill=gray!10, draw=gray!65},
    petsc/.style={layer, fill=red!7, draw=orange!70!black},
    runtime/.style={layer, fill=gray!8, draw=gray!55},
    side/.style={font=\scriptsize, text=black, align=left},
    flow/.style={->, thick, black}
]

\node[top]     (L1) at (0,0.0) {%
    \textbf{JetSCI application layer}\\[-1pt]
    {\scriptsize Newton iteration, nonlinear solve logic, problem setup}
};

\node[jax]     (L2) at (0,-1.18) {%
    \textbf{JAX / \texttt{fea-in-jax}}\\[-1pt]
    {\scriptsize Residual evaluation, \texttt{jacfwd}, \texttt{vmap}, local Jacobians, sparse COO assembly}
};

\node[interop] (L3) at (0,-2.36) {%
    \textbf{CuPy + DLPack}\\[-1pt]
    {\scriptsize Device-array views and GPU-buffer interchange}
};

\node[bridge]  (L4) at (0,-3.54) {%
    \textbf{\texttt{ctypes} + \texttt{petsc4py} + \texttt{mpi4py}}\\[-1pt]
    {\scriptsize Python/C bridge, PETSc object construction, rank-local assembly}
};

\node[petsc]   (L5) at (0,-4.72) {%
    \textbf{PETSc}\\[-1pt]
    {\scriptsize \texttt{Mat}, \texttt{Vec}, \texttt{KSP}, \texttt{PC}, distributed sparse solve}
};

\node[runtime] (L6) at (0,-5.90) {%
    \textbf{MPI + GPU runtime}\\[-1pt]
    {\scriptsize Inter-node communication and device execution}
};

\draw[flow] (L1.south) -- (L2.north);
\draw[flow] (L2.south) -- (L3.north);
\draw[flow] (L3.south) -- (L4.north);
\draw[flow] (L4.south) -- (L5.north);
\draw[flow] (L5.south) -- (L6.north);

\coordinate (r1) at ($(L5.east)+(0.55,0)$);
\coordinate (r2) at ($(r1 |- L1.north)+(0,0.28)$);
\coordinate (r3) at ($(L1.north)+(0,0.28)$);
\draw[flow] (L5.east) -- (r1) |- (r2) -- (r3) -- (L1.north);
\node[side, anchor=west] at ($(r1)+(0.10,1.65)$) {Update\\returned\\to JAX};

\end{tikzpicture}
\caption{JetSCI software stack. JAX / \texttt{fea-in-jax} handles local differentiated finite-element kernels and Newton iteration, CuPy and DLPack provide device-array interoperability, \texttt{ctypes} together with \texttt{petsc4py} and \texttt{mpi4py} bridges to distributed PETSc objects, and PETSc performs the sparse linear solve.}
\label{fig:jetsci_stack}
\end{figure}
JetSCI's implementation is naturally divided into two phases: construction and solution. These phases align closely with the strengths of JAX and PETSc, respectively, and are connected by a no-copy data-transfer pipeline. The full software stack is shown in Figure~\ref{fig:jetsci_stack}.

\paragraph{Construction} After the problem specification is defined --- including boundary conditions, material properties, meshes, and indexing --- this information is passed to JAX for construction of the finite-element system. The domain is partitioned into batches of similar materials in order to maximize the amount of computation that can be \texttt{jit}-compiled and \texttt{vmap}-vectorized, since JAX performs best when array shapes and execution patterns are consistent across a batch.

Because the solve requires a Jacobian, JetSCI constructs it at the batch level. Each batch contains a set of elements of the same type, a map from local degrees of freedom to global matrix entries, and a residual kernel determined by the material model and constitutive law. For each batch, JAX automatically differentiates the residual kernel and vectorizes this operation across the batch, producing the dense local Jacobian contributions in parallel. These local Jacobians are then collected and unrolled, and their positions in the global sparse Jacobian are determined from the batch indexing information. The resulting values and indices are assembled into a sparse tuple representation and passed through a buffer callback.

\paragraph{Solution} This marks the start of the solution phase. The buffer callback allows JetSCI to avoid forcing JAX to stage data through host memory. However, two interface limitations must be addressed: JAX arrays do not directly expose the underlying device pointer, and \texttt{petsc4py} does not natively permit matrix construction from user-supplied device pointers. JetSCI resolves this by using DLPack, CuPy, and \texttt{ctypes}. First, the JAX data are converted through DLPack into a CuPy array. This conversion avoids copying, since the DLPack interface is specifically designed for zero-copy-style tensor exchange. CuPy then exposes the device pointer associated with the data. Although PETSc's native C interface can construct matrices and vectors directly from such pointers, \texttt{petsc4py} does not expose this path directly. JetSCI therefore uses \texttt{ctypes} to call the underlying PETSc matrix-construction routines referenced by \texttt{petsc4py}. The extracted device pointers are passed as arguments to these PETSc routines, yielding PETSc-backed matrix objects without CPU staging.

Once the linear solve is completed in PETSc, the solution is converted through DLPack back into a CuPy array and placed into the callback buffer. Upon completion of the callback, this data are returned to JAX, where the Newton iteration continues.

This design enables JetSCI to use PETSc's solver infrastructure with very low transfer overhead while preserving the JAX-based finite-element workflow, including \texttt{jit}-compilation. Moreover, once the data are represented as PETSc objects, they become naturally amenable to MPI-distributed execution, allowing PETSc to distribute matrices and vectors across ranks for scalable parallel solution.
\section{Results}
\label{sec:results}
\subsection{Experimental Methodology}
\label{sec:exp-meth}

A strictly one-to-one comparison between pure-JAX and mixed JAX--PETSc workflows is not always meaningful, since the purpose of JetSCI is precisely to expose solver and parallel capabilities that are not naturally available in a JAX-only implementation. We therefore evaluate the methods componentwise. In particular, we compare Jacobian construction strategies, linear-solve strategies, and scaling behavior across single-node and multi-GPU settings. The goal is not to force identical solver stacks, but to compare the most natural realizations of the same finite-element workflow under different software backends.

As a benchmark, we consider a heterogeneous two-material finite-element problem motivated by composite micromechanics. This benchmark is inspired by the class of fiber/matrix microstructure problems studied by Ballard, McLendon, and Whitcomb\cite{ballard2014microstructure}, who considered finite-element models of random composite microstructures in order to infer fiber elastic properties from lamina- and matrix-level data. In that setting, the essential computational feature is the same one that drives JetSCI: the domain consists of distinct material phases governed by different constitutive behavior, so that the global residual and Jacobian are assembled from phase-dependent local element contributions\cite{ballard2014microstructure}. Our benchmark captures this heterogeneous two-phase structure, although we do not reproduce the full inverse-identification procedure of that work. 

Concretely, the computational domain contains two material regions with distinct constitutive models. For each Newton iteration on the residual system, each phase contributes its own local residual and Jacobian blocks, and these local contributions are assembled into a single sparse global Jacobian that varies per Newton iteration. We then solve the resulting linearized system with a (iterative) Krylov solver as one step of the overall nonlinear iteration. This setup provides a controlled way to compare local differentiated assembly against global sparse solution, which is the central software boundary in JetSCI.

We consider three meshes of increasing resolution. As the mesh is refined, the number of degrees of freedom and the size of the assembled Jacobian increase accordingly. For the JAX-only implementation, we study both a matrix-free variant based on Jacobian--vector products and a matrix-explicit variant in which the Jacobian is formed directly. For JetSCI, the Jacobian is formed explicitly in JAX, transferred into PETSc, and solved there using PETSc's sparse linear algebra infrastructure. These tests therefore separate the effect of local differentiated assembly from the effect of the global solver backend.

To study both performance and scalability, we ran the experiments in two environments: a single-node workstation and a multi-GPU cluster setting. All single-node experiments were conducted on a workstation running Ubuntu 24.04.4 LTS, equipped with an AMD Ryzen 9 7950X processor (16 cores, 32 threads, up to 5.88\,GHz), an NVIDIA GeForce RTX 4080 GPU with 16\,GB of VRAM, and 64\,GB of system RAM. The CUDA toolkit version was 13.2.

Multi-GPU experiments were performed on the Notchpeak cluster at the University of Utah Center for High Performance Computing (CHPC). Each node used in our experiments is equipped with two Intel Xeon Gold 6230 processors (20 cores per socket, 40 cores and 80 threads total, up to 3.9\,GHz), 376\,GB of system RAM, and an InfiniBand interconnect. All cluster experiments were allocated on a single node with two NVIDIA RTX A5500 GPUs and 64\,GB of GPU-accessible memory via the \texttt{notchpeak-gpu-guest} partition. The CUDA toolkit version on Notchpeak was 12.8.

Both systems used Python 3.12 and PETSc 3.24. The workstation used OpenMPI 4.1.6 and PETSc 3.24.5, while Notchpeak used OpenMPI 5.0.8 and PETSc 3.24.1. All experiments used the latest available JAX release at the time of testing.

\subsection{Timings}
We now present timings on various experiments comparing JetSCI to a JAX-only setting. In all figures, a line that terminates before reaching the largest problem size indicates solver divergence at that scale. All experiments used a single GPU unless otherwise indicated.\\
\textbf{Sparse Matrix-vector Products:}
\begin{figure}[!htpb]
\centering
\includegraphics[width=0.75\linewidth]{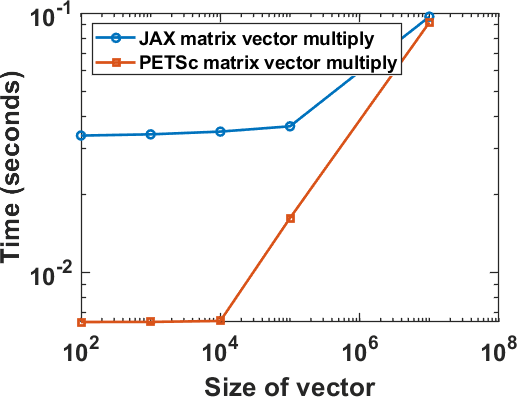}
\caption{Comparison of time to complete 100 matrix-vector products in both PETSc and JAX on a single GPU.}
\label{fig:MatVecCost}
\end{figure}
The fundamental motivation for JetSCI's design is that PETSc's sparse matrix-vector products (used in Krylov solvers) outperform their JAX equivalents. This is demonstrated directly in the following plot, which establishes that the per-iteration cost of the global solve favors PETSc across all tested problem sizes. Given this, JetSCI's overall performance is governed primarily by setup and transfer overhead rather than solve time; the no-copy GPU transfer pipeline is designed specifically to minimize this overhead, so that PETSc's per-iteration advantage is not eroded by data movement costs. \\
\textbf{Jacobi-preconditioned Iterative Solves}:
\begin{figure}[!htpb]
\centering
\includegraphics[width=0.75\linewidth]{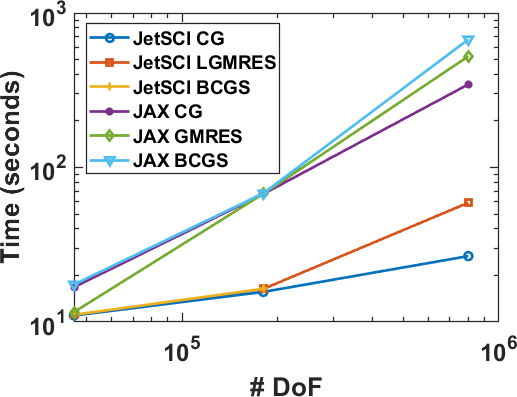}
\caption{Comparison of time to convergence between PETSc and JAX solvers on a single GPU with Jacobi preconditioning.}
\label{fig:JacobiJaxPETSc}
\end{figure}
Figure~\ref{fig:JacobiJaxPETSc} compares JAX and JetSCI iterative solvers (as enabled by PETSc) in double precision on the same explicitly formed sparse matrix, using CG, GMRES, and BCGS with Jacobi preconditioning, timed to reach a relative residual of $1e-13$. JetSCI solvers outperform their JAX counterparts at all tested problem sizes and are more robust, with JAX solvers diverging at larger scales where JetSCI continues to converge.
\\
\textbf{JAX-only Baselines}:
\begin{figure}[!htpb]
\centering
\includegraphics[width=0.75\linewidth]{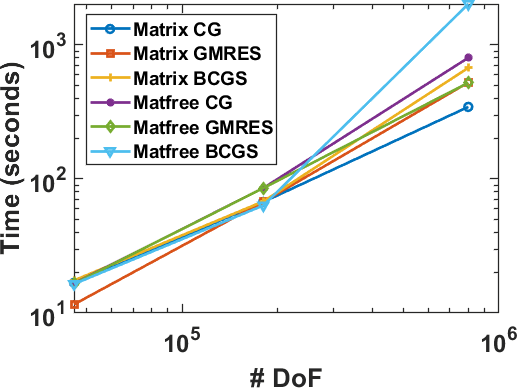}
\caption{Comparison of time to converge for JAX-only matrix-explicit and matrix-free solvers.}
\label{fig:MatVsMatfree}
\end{figure}
For completeness of baseline, Figure~\ref{fig:MatVsMatfree} compares the matrix-explicit and matrix-free approaches using a JAX-only framework, without PETSc integration. The results show no major differences between the matrix-based and matrix-free solvers as a whole. However, comparing Figures \ref{fig:JacobiJaxPETSc} and \ref{fig:MatVsMatfree}, it is clear that JetSCI (enabled by PETSc) outperforms JAX-only variants, regardless of the paradigm used in JAX (matrix-explicit or matrix-free). 
\\
\textbf{Direct Solvers}: 
\begin{figure}[!htpb]
\centering
\includegraphics[width=0.75\linewidth]{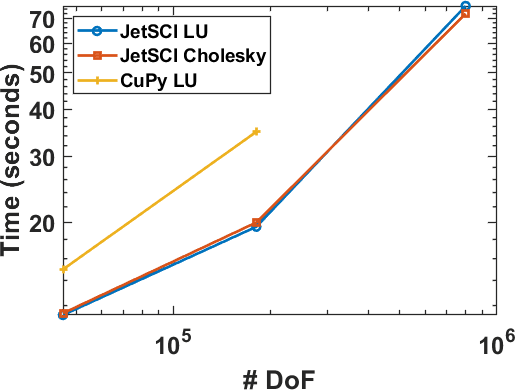}
\caption{Convergence time for direct solvers in CuPy and JetSCI.}
\label{fig:DirectSolvers}
\end{figure}
Sparse direct solvers offer an attractive alternative to iterative solvers for a wide range of problems. Figure~\ref{fig:DirectSolvers} compares direct solvers in JetSci (as enabled by PETsc) against those enabled by CuPy. CuPy's \texttt{spsolve} function, which calls SuperLU\cite{demmel1999supernodal}, is compared against PETSc LU and Cholesky factorizations; while the latter are typically invoked as preconditioners, PETSc allows for their use as direct solvers also. PETSc's direct solvers (used by JetSCI) converge faster and remain stable at larger problem sizes where CuPy's SuperLU implementation begins to struggle.
\\
\textbf{Effect of Preconditioning}: 
\begin{figure}[!h]
\centering
\includegraphics[width=0.75\linewidth]{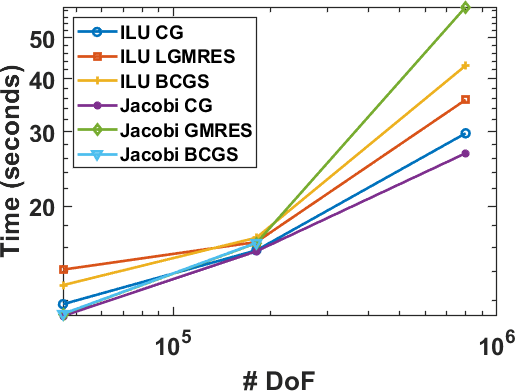}
\caption{Convergence time for JetSCI solvers with Jacobi and ILU preconditioning.}
\label{fig:ILUJacobi}
\end{figure}
Figure~\ref{fig:ILUJacobi} shows the effect of preconditioner choice on JetSCI's iterative solvers, comparing Jacobi and ILU preconditioning across CG, GMRES, and BCGS. ILU preconditioning provides a consistent reduction in convergence time (except for CG, where the two are comparable on this problem), particularly at larger problem sizes where the condition number of the assembled Jacobian increases. This comparison is only possible because JetSCI constructs an explicit sparse matrix; ILU factorization cannot generally be applied to matrix-free operators.
\\
\textbf{Multi-GPU experiments}: 
\begin{figure}[!htpb]
\centering
\includegraphics[width=0.75\linewidth]{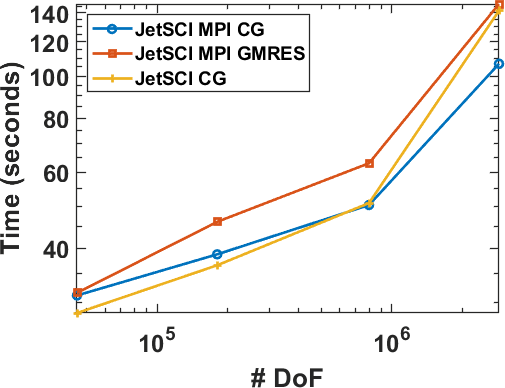}
\caption{Convergence time for JetSCI on single GPUs vs multiple GPUs on Notchpeak.}
\label{fig:MPI}
\end{figure}
Finally, Figure~\ref{fig:MPI} shows the effect of distributing the linear solve across multiple GPUs using MPI on the Notchpeak cluster. Because the RTX A5500 GPUs available on the cluster are older and less computationally powerful than the RTX 4080 used in the single-node experiments, absolute timings are not directly comparable across the two environments; the single-GPU baseline in this plot reflects the A5500 to keep comparison fair. At small problem sizes, the MPI communication overhead slightly increases wall time relative to the single-GPU case, but as problem size grows this overhead is overtaken by the benefit of distributing memory and computation across multiple devices. The crossover reflects the point at which the problem no longer fits comfortably within the fast memory of a single GPU.
\section{Conclusion}
We presented JetSCI, a hybrid JAX--PETSc framework for differentiable finite-element simulation that combines JAX's strengths in local differentiated kernel construction with PETSc's strengths in large-scale sparse algebraic solution. Across the heterogeneous micromechanics problems considered here, this combination was consistently advantageous: JetSCI outperformed JAX-only solver paths in efficiency, robustness, and scalability, while simultaneously exposing a far broader solver and preconditioning ecosystem than is available in a pure-JAX workflow. The results therefore support the central claim of this work: for nonlinear finite-element problems with complex local physics and large sparse global systems, it is neither necessary nor desirable to force the entire simulation stack into a single differentiable runtime. A more effective design is to differentiate where differentiation is most natural, and to solve where large-scale sparse solvers are strongest.

More broadly, JetSCI shows that modern differentiable programming frameworks and mature HPC solver libraries are not necessarily competing paradigms, but can be complementary layers of the same computational stack. JAX provides a productive and accelerator-friendly environment for residual evaluation, constitutive modeling, and Jacobian generation; PETSc provides the distributed-memory infrastructure, solver flexibility, and preconditioning technology needed once those local contributions have been assembled into a global system. By connecting these layers through a low-overhead GPU-resident transfer pipeline, JetSCI preserves the benefits of both without collapsing to the limitations of either.

There remains substantial room for future development. In the current implementation, MPI-enabled PETSc objects can be distributed efficiently once constructed, but matrix and vector generation are still more constrained than the downstream distributed solve. Extending the construction phase to better exploit distributed-memory parallelism by introducing domain decomposition into the JAX workflow, reducing remaining interop overheads, and broadening support for additional nonlinear and multiphysics workflows are natural next steps. Even in its present form, however, JetSCI already demonstrates a clear practical lesson: hybrid differentiable-HPC frameworks can unlock problem sizes, solver strategies, and levels of performance that are difficult to access from JAX-only approaches. We expect this style of design to become increasingly important as scientific machine learning methods are pushed toward the scale, complexity, and robustness requirements of production simulation.

\section*{Acknowledgments}
The authors would like to acknowledge and thank the High Performance Computing Modernization Program (HPCMP), the HPC Internship Program (HIP) sponsorship, and the Oak Ridge Institute for Science \& Education (ORISE). Furthermore, we also thank the HPCMP PET program for making this framework possible. The authors would also like to acknowledge that ChatGPT was used to correct grammar throughout the paper

\bibliographystyle{IEEEtran}
\bibliography{IEEEabrv,references}








\end{document}